\documentstyle[ichep,12pt]{article}
\newif\ifpreprint
\preprinttrue

\newcommand{\beq}{\begin{equation}}
\newcommand{\eeq}{\end{equation}}

\def\ra{\rightarrow}
\def\x{\times}

\def\beqa{\begin{eqnarray}}
\def\eeqa{\end{eqnarray}}

\title{SIGNATURES FOR HEAVY $\bf Z'$ BOSONS AT HADRON
COLLIDERS\ifpreprint\thanks{Invited Talk given at the
XXVI International Conference on High Energy Physics, August 6--12,
1992, Dallas, Texas}\fi}

\author{
Mirjam Cveti\v{c}\ifpreprint\thanks{email:
cvetic@cvetic.hep.upenn.edu}\fi\ \
 and Paul Langacker\ifpreprint\thanks{email:
pgl@langacker.hep.upenn.edu}\fi \\ University of
Pennsylvania \\ Department of Physics \\ Philadelphia, Pennsylvania,
USA  19104-6396}

\begin{document}
\finalcopy
\ifpreprint\pagestyle{myheadings}
\markright{\vbox{\rm\noindent UPR--533--T\hfill\break\vskip 3mm
\noindent September, 1992\hfill\vskip 4mm}}\fi
\maketitle
\abstract{
  We discuss three promising diagnostic probes
of the couplings of a possible heavy $Z'$ boson at hadron colliders:
the forward-backward asymmetry into leptons, the rare decay $Z' \ra
W^{\pm} \ell^{\mp} \nu$, and associated production $pp \ra Z'W,\;
Z'Z$.  A general model-independent framework is proposed for
interpreting these and other data.  Clean tests of whether a new $W'$
couples to right-handed currents, of the ratio $g_R/g_L$ of gauge
couplings, and of the non-abelian vertex in left-right symmetric
models are described.}


Many extensions of the standard model (SM), including string theories,
predict the existence of an
extra $U_1$ gauge symmetry and the associated $Z'$ boson.
In most cases there is no
prediction for the mass of the new particles,
 but it is at
least plausible that they could be
in the accessible TeV range.
(Admittedly, this may be a case of looking under the lamppost for the
new physics.)

Existing limits from $Z$-pole, neutral current, and collider data give
stringent limits on $Z-Z'$ mixing~\cite{r1}--\cite{LLM}.
However, the limits on  the
actual masses of new $Z'$ bosons are relatively weak, typically in the
range 160--400~GeV depending on the model.  This is sometimes
strengthened to 500--1000~GeV in specific models in which the mass and
mixing are correlated.

If a new $Z'$ is discovered either via indirect
\ifpreprint\newpage\noindent\fi
precision tests or by
direct production at new
colliders then the next stage will be to try to
determine its properties.
In particular, one will want to
determine the charges
 $\hat{g}^{f}_{L,R}$, which describe the $Z'$
couplings
to left and right-handed fermion $f$, and the
overall strength of the gauge coupling $g_2$~\cite{RR,C1,C2}.
If a new charged $W'$ is discovered one will also want to be
able to probe the nature of its couplings to fermions.

Heavy $Z'$ or $W'$
bosons can be discovered
at hadron colliders~\cite{EE} by their leptonic decays
for masses up to $\sim 5$~TeV~\cite{LRR,BAR,AG,HR,r1}, assuming
couplings as expected in typical grand unified theories.
The primary goal                following the discovery of a heavy $Z'$
would be to
       determine the nature of the gauge group, {\it i.e.,} to determine
the normalized interactions of ordinary fermions.

The neutral current gauge interaction term in the presence of an
additional $U_1$ can be written as:
\beq -L_{NC}=eJ_{em}^\mu A_\mu +g_1 J_1^\mu Z_{1\mu}+ g_2J_2^\mu
Z_{2\mu}, \label{EQNC} \eeq
with $Z_1$ being the $SU_2 \times U_1$ boson and $Z_2$ the additional
boson in the weak eigenstate basis.  Here $g_1 \equiv
\sqrt{g_L^2+g_Y^2}=g_L/\cos\theta_W$, where $g_L$, $g_Y$ are the gauge
couplings of $SU_{2L}$ and $U_{1Y}$, and $g_2$ is the gauge coupling
of $Z_2$.  The GUT-motivated cases have $g_2=\sqrt{5/3}\sin\theta_W
g_1\lambda_g^{1/2}$, where $\lambda_g$ depends on the symmetry
breaking pattern \cite{RR}.  If the GUT group breaks directly to
$SU_3\times SU_2\times U_1\times U_1'$, then $\lambda_g=1$.

The currents in (\ref{EQNC})  are:
\beq J_j^\mu=\frac{1}{2}\sum_i{\bar{\psi}_i}\gamma^\mu
\left[\hat{g}^i_{Vj}-\hat{g}^i_{Aj}\gamma_5\right] \psi_i, \quad
j=1,2, \label{CURR}\eeq
where the sum runs over fermions, and the $\hat{g}^i_{(V,A)j}$
correspond to the vector and axial vector couplings of $Z_j$ to the
$i^{th}$ flavor.  Analogously, $\hat{g}^i_{(L,R)j}=\frac{1}{2}(
\hat{g}^i_{Vj}\pm \hat{g}^i_{Aj})$, so that $\hat{g}_{L1}^i= t_{3L}^i -
\sin^2 \theta_W q^i$ and $\hat{g}_{R1}^i = - \sin^2 \theta_W q^i$, where
$t_{3L}^i$ and $q^i$ are respectively the third component of weak
isospin and electric charge of fermion $i$.

We will concentrate on extra $U_1$'s which occur in grand unified
theories, such as in the breakings
$SO_{10} \ra SU_5 \x U_{1\chi}$, $E_6 \ra SO_{10} \x
U_{1\psi}$, or $E_6 \ra SU_3 \x SU_2 \x U_1 \x U_{1\eta}$, where
$Q_\eta = \sqrt{\frac{3}{8}} Q_\chi -
\sqrt{\frac{5}{8}} Q_\psi$. We will also consider left-right
symmetric models based on the $SU_3 \x SU_{2L} \x SU_{2R} \x U_{1
B-L}$ \cite{C2}.
The charges for these models are listed
in \cite{r2}.
In the following we will assume family universality and  ignore possible
$Z - Z'$ mixing.

The          fermion charges identify the model or the nature of the
group.            The overall normalization is arbitrary so it is
actually the ratios of the normalized charges $\hat{g}^f_{L,R2}$ that are
relevant.  Unfortunately, one cannot determine the            charges
from the production rate for charged leptons \cite{C3}, $\sigma (pp \ra
Z') B(Z' \ra \ell^+\ell^-)$,          because the branching ratio
depends on
the total width $\Gamma_{Z'}$.
In most models there are possible exotic fermions and supersymmetric
partners           which can lead to  a model dependence of at least
a factor of two in $\Gamma_{Z'}$.
                                                     Similarly, the
overall production rate is proportional to
$g_2$ which,    even for a given group, depends         on
the symmetry breaking pattern.
We therefore concentrate       on
further tests which determine the ratios of the   normalized charges
and are thus free from these ambiguities.

It is convenient to define the ratios
\beq \gamma^f_L \equiv \frac{\left( \hat{g}^f_{L2} \right)^2}{\left(
\hat{g}^\ell_{L2} \right)^2 + \left( \hat{g}^\ell_{R2} \right)^2}
\;\;\;\;\;
     \gamma^f_R \equiv \frac{\left( \hat{g}^f_{R2} \right)^2}{\left(
\hat{g}^\ell_{L2} \right)^2 + \left( \hat{g}^\ell_{R2} \right)^2} \eeq
of the squares of the left and right chiral couplings to fermion $f$
relative to the total interaction strength to leptons.  We
will further assume that the generators associated with the new $Z'$
commute with the standard model generators, as in the
             $SU_2 \x U_1 \x U'_1$ group or left-right
models~\cite{C3,ZWW2}.  In this case one has automatically that
$\gamma_L^u = \gamma_L^d \equiv \gamma_L^q$, since the $u_L$ and $d_L$
                   are in the same $SU_2$ multiplet, and similarly
$\gamma_L^e = \gamma_L^\nu \equiv \gamma_L^l$.  We also introduce
\beq \tilde{U}\equiv\gamma_R^u/\gamma_L^q \label{tilu} \;\;\;\;
     \tilde{D}\equiv\gamma_R^d/\gamma_L^q. \label{tild} \eeq
The values of $\gamma_L^\ell$, $\gamma_L^q$, $\gamma_R^u$,
$\gamma_R^d$, $\tilde{U}$, and $\tilde{D}$ for the $\chi$, $\psi$,
$\eta$, and $LR$ models are listed in Table~\ref{tabI}.  It is seen
that they vary significantly from model to model.

\begin{table}  \centering
\begin{tabular}{|c|cccc|}            \hline
\ & $\chi$ & $\psi$ & $\eta$ & $LR$ \\ \hline
$\gamma^\ell_L$ & 0.9 & 0.5 & 0.2 & 0.36 \\
$\gamma^q_L$ & 0.1 & 0.5 & 0.8 & 0.04 \\
$\gamma_R^u$ & 0.1 & 0.5 & 0.8 & 1.4 \\
$\gamma_R^d$ & 0.9 & 0.5 & 0.2 & 2.6 \\
$\tilde{U}$ & 1 & 1 & 1 & 37 \\
$\tilde{D}$ & 9 & 1 & 0.25 & 65 \\
$2\tilde{U}+\tilde{D}$ & 11 & 3 & 2.3 & 139 \\ \hline
\end{tabular}
\caption{ Values of $\gamma_L^\ell$, $\gamma_L^q$, $\gamma_R^u$,
$\gamma_R^d$, $\tilde{U}$, $\tilde{D}$, and $2\tilde{U}+\tilde{D}$ for
the $\chi$, $\psi$, $\eta$, and $LR$ ($\kappa=1$) models.}
\label{tabI}
\end{table}

Now let us describe possible clean probes of the ratios
$\gamma^f_{L,R}$                       at hadron colliders.
The first is the forward-backward asymmetry~\cite{LRR}.  For
$pp\rightarrow Z'\rightarrow\ell^+\ell^-$,
\beq
A_{FB}=\frac{\left[\int_0^{y_{max}}-\int_{y_{min}}^0\right]\left[
F(y)-B(y) \right]\,dy}{\int_{y_{\rm min}}^{y_{\rm max}}\left[
F(y)+B(y)\right]\,dy}, \label{EQAFB} \eeq
where $F(y)\pm B(y)=[\int_0^1\pm\int_{-1}^0] \,d\cos\theta (d^2\sigma/
dy\, d\cos\theta)$ and $\theta$ is the $\ell^-$ angle in the $Z'$ rest
frame.
                               $A_{FB}$ can be be expressed in terms
of gauge couplings as:
\beq A_{FB}=\frac{3}{4}\left(2\gamma_L^\ell-1\right) 0.58\frac{1
-0.75\tilde{U}-0.25\tilde{D}}{1 +0.68\tilde{U}+0.32\tilde{D}}
\label{EQAFBP} \eeq
for a 1 TeV $Z'$ at the LHC.
$A_{FB}$ thus depends on $\gamma_L^\ell$, and
on the combination $\simeq2\tilde{U}+\tilde{D}$ of right-handed
currents. $A_{FB}$ is listed for various models in Table~\ref{tabII},
and is shown as a function of $M_{Z'}$ in \cite{r2}, along with the
analogous $A_{FB}$ for heavy $W^\pm_R$ production.  The statistical
errors are small enough to discriminate effectively up to $\simeq2$
TeV, provided that good enough lepton charge identification can be
achieved. $A_{FB}$ would be an excellent determinant of
$\kappa = g_R/g_L$ for
the LR model \cite{C2}.  Typically, one should be able to determine
$2\tilde{U}+\tilde{D}$ to (5 -- 10)\% for $M_{Z'}\simeq1$~TeV if
$\gamma_L^\ell$ is known independently from $r_{\ell\nu W}$ (below).

The forward-backward asymmetry for $pp\rightarrow
W^{\prime\pm}\rightarrow\ell^\pm n_\ell$ does not distinguish $V+A$
couplings of the $W'$ from $V-A$.  As described below, this
information can be obtained from $pp\rightarrow W'Z$ and the
nonobservation of $pp\rightarrow W' W$ and $W'\rightarrow
W\ell^+\ell^-$.

\begin{table*}[t]           
\begin{center}
\begin{tabular}{|ccccc|}            \hline
\ & $Z_\chi$ & $Z_\psi$ & $Z_\eta$ & $Z_{LR}$ \\
\hline
$R_{Z'Z}$ & $\ 0.0021\pm 0.0002$ & $ 0.0046\pm 0.0008$& $ 0.0050\pm
0.0007 $& $ 0.0010\pm 0.0001$
\\ \hline
$R_{Z'W}$ &$   0.0057\pm 0.0004
$& $   0.012\pm 0.001$&$  0.014\pm 0.001 $&
$   0.0005\pm 0.0001$
\\ \hline
$r_{\ell\nu W}$ &$   0.055\pm 0.0014 $& $ 0.030\pm 0.002
$&$ 0.012\pm 0.001 $& $  0.022\pm 0.0008 $
\\ \hline
$A_{FB}^{e^+e^-}$ & $-0.134\pm0.007$ & $0.000\pm0.016$ &
  $-0.025\pm0.014$ & $0.100\pm0.006$
\\ \hline
\end{tabular}
\end{center}
\caption{
Ratios $R_{Z'V}$, $r_{\ell\nu W}$, and $A_{FB}$ with their error bars
at the LHC for $M_{Z'}=1 $ TeV, assuming a
$10^7$s run at a luminosity of $10^{34}$cm$^{-2}s^{-1}$.  The
statistical error bars are slightly larger for the SSC with a
luminosity of $10^{33}$cm$^{-2}s^{-1}$.}
\label{tabII}
\end{table*}

A second useful probe are the rare decays $Z' \ra f_1 \bar{f}_2
V$~\cite{RIZ,C3,ZWW2,AG2}, where $V = W$ or $Z$.  We always sum over
$\ell = e,\mu$; over $W^+,W^-$; and over the neutrino flavors.  The
ratios $r_{f_1 f_2V}\equiv B(Z'\rightarrow f_1 \bar{f}_2
V)/B(Z'\rightarrow\ell^+\ell^-)$ project out different combinations of
the normalized couplings.
For example,
\beq \frac{r_{ffZ}}{a_Z}=\frac{(\hat{g}^f_{L1})^2 (\hat{g}^f_{L2})^2 +
(\hat{g}_{R1}^f)^2 (\hat{g}^f_{R2})^2}{ (\hat{g}^{\ell}_{L2})^2 +
(\hat{g}^{\ell}_{R2})^2}.\label{rffza} \eeq
$a_Z(a_W)$ are
kinematic factors.  For $M_{Z'}=1$ TeV, $a_Z$ ($a_W$) $=0.068$
($0.080$).
These quantities in principle
determine $\gamma_L^\ell$, $\gamma_L^q$, and
$\gamma_R^u+\frac{3}{8}\gamma_R^d$.  In practice, however, the
dependence on the $\gamma_R$'s is very weak.

In hadron colliders the hadronic ratios $r_{had\,Z}$ and $r_{had\,W}$
suffer~\cite{C3} from serious QCD backgrounds.  Similarly, $r_{\nu\nu
Z}$ suffers from the standard model background $pp\ra ZZ$, with one
$Z\ra \nu\bar{\nu}$.  These are probably unobservable at hadron
colliders.

On the other hand, $r_{\ell\ell Z}$ provides a very clean signal.
Unfortunately, it has a weak dependence~\cite{C3} on $\gamma_L^\ell$
due to the fact that $|\hat{g}^\ell_{L1}|\simeq |\hat{g}^\ell_{R1}|$
for $\sin\theta_W^2 \simeq0.23$, and thus $r_{\ell\ell Z}$ serves only
as a consistency check.

The backgrounds for the ratio $r_{\ell\nu W}$ can be eliminated with
appropriate cuts.  The mode with $W$ decaying hadronically can be
separated from the standard model back\-ground\footnote{This assumes that
the $W$ can be reconstructed in events tagged by
a hard lepton.} from $pp\rightarrow WW$
(as well as the events due to $Z-Z'$ mixing) by requiring~\cite{C3} the
$\ell\nu$ transverse mass $m_{T\ell\nu}\geq90$ GeV.  This cut reduces the
signal only by $4\%$ for $M_{Z'}=1$ TeV at the LHC.  Events with
$W\rightarrow\ell\nu$ may also be observable with appropriate
cuts~\cite{AG2}. The ratio
\beq \frac{r_{\ell\nu W}}{a_W} = 0.77 \gamma_L^{\ell} \eeq
thus provides the
``gold-plated'' signal which tests the left-handed coupling of leptons
$\gamma_L^\ell$.  The predictions                     for various
models and typical statistical errors
are listed in Table~\ref{tabII}.  One expects an accuracy of (2-10)\%
on $\gamma^\ell_L$            for the specific models considered here.
$r_{\ell \nu W}$ is also an excellent measure of $\kappa$ in $LR$
models~\cite{C3}.  Finally, in left-right models one does not expect
$W_R \ra W_L \ell^+ \ell^-$ except from very small mixing and fermion
mass effects.  Assuming that the forward-backward asymmetry for a $W'$
is consistent with $V\pm A$, the {\it absence} of $W' \ra W \ell^+
\ell^-$ events would be a clear signal for $V+A$.

The third probe of $Z'$ charges is associated $Z'V$
production~\cite{r2}, which is  sensitive to the quark charges.
We define the ratios
\beq R_{Z'V}=\frac{\sigma (pp\ra Z'V)B(Z'\ra \ell^+\ell^-)}{ \sigma
(pp\ra Z')B(Z'\ra \ell^+\ell^-)}, \label{ratzpv} \eeq
with $V=Z$ or $W$ decaying into leptons or quarks.  We define the
cross section for $pp\ra Z'W$ as the sum over $W^+$ and $W^-$.  In the
models with heavy charged gauge bosons the ratios
\beq R_{W'V}=\frac{\sigma (pp\ra W'V)B(W'\ra \ell{\bar{n}_\ell}+
\bar{\ell} n_\ell)}{\sigma (pp\ra W')B(W'\ra \ell{\bar n_\ell}
+\bar{\ell} n_\ell)}\label{ratwpv} \eeq
can be defined analogously.  The branching ratios involve decay modes
with charged leptons, which provide clean signals and   are free of
major backgrounds, especially for $Z'$.

One has
\beqa R_{Z'Z}&=&10^{-3} \frac{7.0+0.85\tilde{U}+0.09\tilde{D}}{
1+0.68\tilde{U}+0.32\tilde{D}} \\
R_{Z'W}&=&10^{-3} \frac{22.2}{ 1+0.68\tilde{U}+0.32\tilde{D}}
\label{RVPV} \eeqa
for a 1 TeV $Z'$ at the LHC.  Typical values and uncertainties are shown
in Table~\ref{tabII}.  Except in models with large $\tilde{U}$ and
$\tilde{D}$ the $R$'s mainly determine the combination
$2\tilde{U}+\tilde{D}$, typically with a precision of $\sim$ (10 --
20)\%.  This is the same quantity that is probed by $A_{FB}$ (for a
known $\gamma_L^\ell$), but this provides an independent measurement.
In the case of models with heavy charged
gauge bosons the ratio $R_{W'V}$ would also yield information on the
coupling of $W'$ to the quarks.  The absence of $W'W$ events would
again be evidence for $V+A$ currents~\cite{r2}, while in the $LR$
model there is a definite prediction for $R_{W'Z}$ that would test both
the non-Abelian  $W'W'Z$ vertex and the $V+A$ nature of the $W'$
coupling~\cite{r2}.

Finally, if
one can measure the product
of the total leptonic rate times the width of the $Z'$, as observed
from the line shape, one can extract the overall strength parameter
$g_2$,
     which probes the grand unified theory breaking pattern.  Also,
the total width itself                                 is a probe of
exotic fermions and supersymmetric partners.

It may be
possible to utilize the $y$-distributions of the production cross
sections and asymmetries~\cite{AG3} and $R_{Z' \gamma}$~\cite{zpgam}
to learn about the
relative couplings of the up    and down quark, and it is
possible that the polarization of the decay $\tau$ products might be
measurable~\cite{CAHN}.
Polarized proton beams would also be very useful~\cite{taxil}.
A serious background study of the backgrounds
and signatures of the various collider modes is in
progress~\cite{AG3}.

            Future high energy $e^+e^-$
colliders would be sensitive to the properties of both real and
virtual $Z'$'s \cite{EE}.  Indirect
indications from                           atomic parity violation and
other future
precision observables~\cite{r2,LLM} are a completely different realm.
If there should be new heavy gauge bosons with $M \sim 1\ TeV$
     there will be many ways of probing their properties at hadron
colliders, in
indirect precision experiments, and in possible
 new $e^+e^-$ colliders.

\end{document}